\documentclass[a4paper,fleqn]{cas-dc}
\usepackage[numbers,sort&compress]{natbib}
\usepackage{hyperref}
\hypersetup{colorlinks=true,linkcolor=blue,citecolor=magenta,filecolor=magenta, urlcolor=cyan}

\newcommand{\be}{\begin{equation}}
\newcommand{\ee}{\end{equation}}
\newcommand{\bes}{\begin{equation*}}
\newcommand{\ees}{\end{equation*}}
\def\bsp#1\esp{\begin{split}#1\end{split}}
\def\bpm{\begin{pmatrix}}
\def\epm{\end{pmatrix}}
\newcommand{\bea}{\begin{eqnarray}}
\newcommand{\eea}{\end{eqnarray}}

\def\gL{g_{\scriptscriptstyle L}}
\def\gR{g_{\scriptscriptstyle R}}
\def\gBL{g_{\scriptscriptstyle B-L}}
\def\gY{g_{\scriptscriptstyle Y}}
\def\vR{v_{\scriptscriptstyle R}}
\def\vL{v_{\scriptscriptstyle L}}
\def\tw{\theta}
\def\phw{\varphi}
\def\lag{{\cal L}}

\begin{document}
\def\floatpagepagefraction{1}
\def\textpagefraction{.001}

\title [mode = title]{Flavour-changing top quark decays in the alternative left-right model}  
\let\printorcid\relax   

\author[1]{Mariana Frank}
\author[2]{Benjamin Fuks}
\author[3]{Sumit K. Garg}
\author[4]{Poulose Poulose}

\address[1]{Department of Physics, Concordia University,
  7141 Sherbrooke St. West, Montreal, Quebec, Canada H4B 1R6}
\address[2]{Sorbonne Universit\'e, CNRS, Laboratoire de Physique Th\'eorique et  Hautes \'Energies, LPTHE, F-75005 Paris, France}
\address[3]{Manipal Centre for Natural Sciences, Manipal Academy of Higher Education, Dr.T.M.A. Pai Planetarium Building, Manipal-576104, Karnataka, India}
\address[4]{Department of Physics, Indian Institute of Technology Guwahati, Assam 781 039, India}

\shortauthors{M.~Frank {\it et~al.}}
\shorttitle{{Flavour-changing top quark decays  in the alternative left-right model}}

\begin{abstract}
  We examine flavour-changing neutral-current decays of the top quark, $t\to q \gamma$, $t \to qZ$, $t \to q H$, and $ t\to q g$ (with $q=u, c$), in the Alternative Left-Right Model, a left right-symmetric model featuring exotic quarks and light bosons. These decays have a very small probability of occurring within the Standard Model, but they can be enhanced in this model through the presence of the exotic states. While associated signals may be detected directly at the LHC, rare decays have the advantage of offering means to probe new particles indirectly, through loop-contributions. We perform a comprehensive analysis of the model's parameter space to demonstrate the possible existence of enhancements in the corresponding branching ratios, of $10^6$ for the branching ratios $\mathcal{B}(t\!\to\! uZ)$ and $\mathcal{B}(t\!\to\! uH)$, and in the range of $10^{2} - 10^{4}$ for the other decays, relative to the Standard Model. We subsequently determine the preferred parameter space regions of the model in terms of potential of being reached in the near future.
\end{abstract}

\begin{keywords}
top quark decays, flavour violation, left-right models
\end{keywords}

\maketitle

%%%%%%%%%%%%%%%%%%%%%%%%%%%%%%%%%%%%%%%%%%%%%%%%%%%%%%%%%%%%%%%%%%%%%%%%%%%%%%%

\section{Introduction}
\label{sec:introduction}
The LHC continues its testing of the Standard Model (SM) while actively searching for physics Beyond the Standard Model (BSM). These searches unfold on varied complementary fronts. On the one hand, they directly look for collider signals of additional particles and interactions, including particularly those indicative of an extended gauge structure. On the other hand, they also probe BSM effects indirectly by scrutinising physical observables associated with rare phenomena in the SM, which may occur more frequently in various BSM scenarios. A prime example of this is the exploration of new top quark interactions via flavour-changing neutral-current (FCNC) processes. The top quark, the heaviest particle in the SM, possesses a mass close to the electroweak scale. Consequently, studying its nature and properties is believed to shed light on the dynamics behind the origin of elementary particle masses and to reveal potentially new characteristics of electroweak and strong interactions. However, since its discovery in 1995~\cite{D0:1995jca,CDF:1995wbb}, the properties of the top quark have consistently aligned with SM predictions, as demonstrated notably at the LHC where top quarks are abundantly produced.

Rare flavour-changing neutral-current decays of the top quark provide an interesting connection between the heaviest elementary particle in the SM and new bosons or fermions typical of BSM scenarios~\cite{TopQuarkWorkingGroup:2013hxj}. In the SM, tree-level FCNC processes involving a neutral boson are absent due to the unitarity of the Cabibbo-Kobayashi-Maskawa (CKM) matrix, which governs flavour mixings. These FCNC processes can only proceed through loop-induced charged-current exchanges of $W$ bosons and SM fermions. In this case, the corresponding amplitudes are proportional to the squared mass of the light quarks running into the loop diagrams, leading to strong suppression: the associated branching ratios are found to lie in the range of $10^{-15}-10^{-12}$~\cite{Mele:1998ag, Aguilar-Saavedra:2002lwv, Aguilar-Saavedra:2004mfd, Zhang:2013xya, Durieux:2014xla}. These branching ratios are potentially modified by the presence of new physics, which could significantly increase them, for example, to $10^{-4} - 10^{-7}$~\cite{Castro:2022qkg, Durieux:2014xla, Larios:2006pb, Barros:2019wxe}. It is typically the case in the quark-singlet model \cite{Diaz-Cruz:1989tem}, two-Higgs-doublet models~\cite{Eilam:1990zc, Atwood:1996vj, Botella:2015hoa, Abbas:2015cua, Baum:2008qm}, the minimal supersymmetric standard model~\cite{Dedes:2014asa, Cao:2007dk, Lopez:1997xv, Eilam:2001dh}, left-right symmetric~\cite{Gaitan:2004by} and supersymmetric~\cite{Frank:2005vd} models, universal~\cite{Dey:2016cve} and warped~\cite{Gao:2013fxa, Diaz-Furlong:2016ril} extra dimensional models, simplified models~\cite{Liu:2021crr, Crivellin:2022fdf}, and models featuring an extended scalar sector~\cite{Banerjee:2018fsx} or anomalous couplings~\cite{Aguilar-Saavedra:2008nuh, Agram:2013koa, Oyulmaz:2018irs}.

Top FCNC couplings have been studied experimentally both through the production of a top-antitop pair followed by a rare decay, and through FCNC single top production followed by a standard decay. To date, no evidence of new effects has been observed, and constraints have consequently been derived from searches in various experiments, including electron-positron collisions at LEP~\cite{L3:2002hbp, OPAL:2001spi, ALEPH:2002wad}, deep inelastic scattering processes at HERA~\cite{ZEUS:2003vfj, ZEUS:2011mya}, and $p {\bar p}$ collisions at the Tevatron \cite{Kikuchi:2000sv, D0:2010dry, D0:2007wfn}. At the LHC, the ATLAS~\cite{ATLAS:2023ujo, ATLAS:2023qzr, ATLAS:2022gzn, ATLAS:2021amo, ATLAS:2018zsq, ATLAS:2018jqi, ATLAS:2018xxe} and CMS~\cite{CMS:2016uzc, CMS:2017twu, CMS:2017wcz, CMS:2021hug, CMS:2023tir, CMS:2020utv} collaborations have significantly improved these bounds by nearly one order of magnitude. The most up to date limits for decays into a $Z$ boson via left-handed (LH) and right-handed (RH) $tZq$ couplings are:
\bes\bsp
  \mathrm{LH:} &\ {\cal B} (t \!\to\! c Z) \le 1.3 \times 10^{-4}\,,\ 
  {\cal B} (t \!\to\! u Z) \le 6.2 \times 10^{-5}\,,\\
  \mathrm{RH:} &\ {\cal B} (t \!\to\! c Z) \le 1.2 \times 10^{-4}\,,\ 
  {\cal B} (t \!\to\! u Z) \le 6.6 \times 10^{-5}\,.
\esp\ees
In addition, the best limits on decays into the SM Higgs boson $H$ are determined from the analysis of potential signals in multi-leptonic and di-photon final states given, for the electron/muon, hadronic tau and di-photon channels, by
\bes\bsp
  e/\mu: &\ {\cal B} (t \!\to\! c H) \le 1.6 \times 10^{-3}\,,\ 
  {\cal B} (t \!\to\! u H) \le 1.9 \times 10^{-3}\,,\\
  \tau: &\ {\cal B} (t \!\to\! c H) \le 9.4 \times 10^{-4}\,,\
  {\cal B} (t \!\to\! u H) \le 6.9 \times 10^{-4}\\
  \gamma: &\ {\cal B} (t \!\to\! c H) \le 5.8 \times 10^{-4}\,,\
  {\cal B} (t \!\to\! u H) \le 1.9 \times 10^{-4}\,.
\esp\ees
Finally, bounds on decays into massless bosons are:
\bes\bsp
 & {\cal B} (t \!\to\! q \gamma) \lesssim \times 10^{-5}\,,\\
 &{\cal B} (t \!\to\! c g) \leq 4.1 \times 10^{-4}\,,\quad 
 {\cal B} (t \!\to\! u g) \lesssim 2 \times 10^{-5}\,.
\esp\ees
Given the prominent role of top quark physics in future collider projects proposed within the community, significant improvements in the aforementioned bounds could be anticipated in case one of such colliders is built~\cite{Mandrik:2018yhe, Zarnecki:2018lup, Cakir:2018ruj, CLICdp:2018esa, TurkCakir:2017rvu, Wang:2017pdg}.

In the present work, we analyse for the first time BSM effects on rare top decays in the framework of Alternative Left-Right-symmetric Models (ALRM)~\cite{Babu:1987kp, Ma:2010us}. Such models emerge from the breaking of an $E_6$ grand-unified symmetry group into its $SU(3)\times SU(3)\times SU(3)$ subgroup, which is next broken into the $SU(3)_c \times SU(2)_L \times SU(2)_H \times U(1)_X$ group that embeds the SM gauge symmetry. While in usual Left-Right Symmetric Models (LRSM)~\cite{Pati:1974yy, Mohapatra:1974gc, Senjanovic:1975rk, Mohapatra:1977mj}, $SU(2)_H$ is identified with $SU(2)_R$ and
$U(1)_X$ with $U(1)_{B-L}$, in the ALRM we associate $SU(2)_H$ with an $SU(2)_{R'}$ group differently embedding the SM fermions into doublets~\cite{Ma:1986we, Frank:2005rb, Ashrythesis, Ashry:2013loa}. In this way, the $SU(2)_{R'}$ partner of the right-handed up-quark $u_R$ is not its down-type counterpart $d_R$ but an exotic down-type quark $d_R'$, and similarly the $SU(2)_{R'}$ partner of the right-handed charged lepton $\ell_R$ is a new neutral lepton, the scotino $n_R$, rather than a more traditional right-handed neutrino $\nu_R$. The $\nu_R$ and $d_R$ states hence remain singlets under both the $SU(2)_L$ and $SU(2)_{R'}$ symmetries, like the LH partners of the new states, $d_L'$ and $n_L$.

ALRM scenarios have several advantages over LRSM ones. In the LRSM, the properties of the Higgs sector could imply non-acceptable tree-level flavour-violating interactions, conflicting with the observed properties of kaon and $B$-meson systems. As a consequence, the $SU(2)_R \times U(1)_{B-L}$ symmetry has to be broken at a very high energy scale, leading to extremely massive extra Higgs and gauge bosons unlikely to be detected at the LHC. In the ALRM, the new scalars always couple simultaneously to ordinary and exotic fermions so that such bounds can be evaded even in the presence of light additional Higgs bosons. In addition, unlike in the LRSM~\cite{Bahrami:2016has}, the ALRM could predict a viable DM candidate as the scotino $n_R$~\cite{Frank:2019nid}.

Previous studies on the ALRM have established conditions for the stability of its vacuum state~\cite{Frank:2021ekj}, shown that the model can have a significant impact on neutrinoless double beta decay processes and leptogenesis~\cite{Frank:2020odd}, and that it could embed a consistent dark matter phenomenology~\cite{Frank:2022tbm}. In this work, we focus on the indirect effects of ALRM exotic fermions and Higgs bosons through their loop-induced contributions to rare top decays. We demonstrate the possibility of significant enhancements over the SM predictions in specific scenarios, with obtained values close to current limits. ALRM scenarios, therefore, have the potential to be tested indirectly in the near future, as bounds on rare top decays are expected to strengthen. In the following, we briefly describe in Sec.~\ref{sec:ALRMmodel} the ALRM before exploring its implications for rare top decays in Sec.~\ref{sec:topraredecay}. We conclude in Sec.~\ref{sec:conclusion}.

%%%%%%%%%%%%%%%%%%%%%%%%%%%%%%%%%%%%%%%%
\section{Alternative Left-Right Models} 
\label{sec:ALRMmodel}
%%%%%%%%%%%%%%%%%%%%%%%%%%%%%%%%%%%%%%%%
In this section we briefly review the ALRM, and refer to~\cite{Frank:2019nid} for a comprehensive introduction. The ALRM is based on the $SU(3)_c \times SU(2)_L \times SU(2)_{R'} \times U(1)_{B-L}$ gauge group, to which we supplement an additional $U(1)_S$ global symmetry. The spontaneous breaking of $SU(2)_{R'} \times U(1)_{B-L} \times U(1)_S$ to the hypercharge group is implemented such that the generalised lepton number $L=S+T_{3R}$ (with $T_{3R}$ being the third generator of $SU(2)_{R'}$ and $S$ the $U(1)_S$ charge) remains unbroken. This is achieved by means of an $SU(2)_{R'}$ doublet of scalar fields $\chi_R$ charged under $U(1)_S$, which we pair with an $SU(2)_L$ counterpart $\chi_L$ to maintain left-right symmetry. This extra $SU(2)_L$ field is, however, blind to $U(1)_S$. The electroweak symmetry is further broken to electromagnetism by means of a  bi-doublet of Higgs fields charged under both $SU(2)_L$ and $SU(2)_{R'}$, but with no $B-L$ quantum numbers. We collect all scalar and fermion fields of the model in Table~\ref{tab:content}, along with the representation of the fields under the model's gauge group and the associated $U(1)_S$ quantum numbers. Associated electric charges are derived through a generalised Gell-Mann-Nishijima relation $Q = T_{3R} + T_{3L} + Y_{B-L}$, where $T_{3L}$ is the third generator of $SU(2)_L$ and $Y_{B-L}$ the $U(1)_{B-L}$ charge.

{\begin{table}
  \setlength\tabcolsep{1.3pt}
  \begin{center}
  \resizebox{\columnwidth}{!}{
  \begin{tabular}{ccc|ccc}
      Field & Repr.&$U(1)_S$ &Fields & Repr.&$U(1)_S$~~ 
      \\\hline &&&&&\\[-.3cm]
      $Q_L = \bpm u_L\\d_L\epm$ &
        $\big({\bf 3}, {\bf 2}, {\bf 1}\big)_{\frac16}$ & 0 &$L_L = \bpm \nu_L\\e_L\epm$ &
        $\big({\bf 1}, {\bf 2}, {\bf 1}\big)_{-\frac12}$ & 1 \\[.3cm]
      $Q_R = \bpm u_R\\d_R'\epm$ &
        $\big({\bf 3}, {\bf 1}, {\bf 2}\big)_{\frac16}$ & $-\frac12$ & $L_R = \bpm n_R\\e_R\epm$  &
        $\big({\bf 1}, {\bf 1}, {\bf 2}\big)_{-\frac12}$ & $\frac32$ \\[.3cm]
        $d'_L$ & $\big({\bf 3}, {\bf 1}, {\bf 1}\big)_{-\frac13}$ & $-1$ & $n_L$   & $\big({\bf 1}, {\bf 1}, {\bf 1}\big)_{0}$ & 2 \\[.1cm]
         $d_R$  & $\big({\bf 3}, {\bf 1}, {\bf 1}\big)_{-\frac13}$ & 0 & $\nu_R$ & $\big({\bf 1}, {\bf 1}, {\bf 1}\big)_{0}$ & 1  \\[.1cm]
     $\chi_L = \bpm\chi_L^+ \\\chi_L^0\epm$ &
         $\big({\bf 1}, {\bf 2}, {\bf 1}\big)_{\frac12}$ & 0& $\chi_R = \bpm\chi_R^+ \\\chi_R^0\epm$ &
         $\big({\bf 1}, {\bf 1}, {\bf 2}\big)_{\frac12}$ &$\frac12$\\[.3cm]
         $\phi = \bpm \phi_1^0&\phi_2^+\\ \phi_1^-&\phi^0_2 \epm$ &
         $\big({\bf 1}, {\bf 2}, {\bf 2}^*\big)_{0}$ & $-\frac12$ &&&\\
    \end{tabular}}
    \caption{ALRM fermions and scalars, their representation under $SU(3)_c\times SU(2)_L\times SU(2)_{R'} \times U(1)_{B-L}$, and $U(1)_S$ quantum numbers.}
    \label{tab:content}
  \end{center}
\end{table}}

In the ALRM, FCNC interactions in the fermionic sector are generated from the Yukawa Lagrangian 
\be\bsp
  & \lag_{\rm Y} = \bar Q_L {\bf \hat Y}^u \hat\phi^\dag Q_R
    - \bar Q_L {\bf \hat Y}^d \chi_L d_R
    - \bar Q_R {\bf \hat Y}^{d'}\chi_R d'_L \\
 &\quad -\bar L_L {\bf \hat Y}^e \phi L_R
    + \bar L_L {\bf \hat Y}^\nu \hat\chi_L^\dag \nu_R
    + \bar L_R {\bf \hat Y}^n \hat\chi_R^\dag n_L +  \mathrm{H.c.}\,,
\esp\ee
in which we omitted all flavour indices for simplicity. The Yukawa couplings ${\bf \hat Y}$ are thus $3 \times 3$ matrices in the flavour space. Moreover, we introduced the dual fields $\hat\phi=\sigma_2\phi\sigma_2$ and $\hat\chi_{L,R}= i \sigma_2 \chi_{L,R}$ with $\sigma_2$ being the second Pauli matrix. The scalar potential dictating Higgs spectrum and mixings includes bilinear ($\mu$), trilinear ($\kappa$) and quartic ($\lambda$, $\alpha$) terms,
\be\bsp
  & \hspace*{-.4cm}V =
    \kappa \big[\chi_L^\dag \phi \chi_R + \chi_R^\dag\phi^\dag\chi_L\big]
   -\mu_1^2 {\rm Tr} \big[\phi^\dag \phi\big]
\\&\hspace*{-.4cm}
  -\mu_2^2 \big[\chi_L^\dag \chi_L + \chi_R^\dag \chi_R\big]
   + \lambda_1 \big({\rm Tr}\big[\phi^\dag \phi\big]\big)^2
\\&\hspace*{-.4cm}
   + \lambda_2\ (\phi\!\cdot\!\hat\phi)\ (\hat\phi^\dag\!\cdot\!\phi^\dag)
   + \lambda_3 \Big[\big(\chi_L^\dag \chi_L\big)^2 + \big(\chi_R^\dag\chi_R\big)^2\Big]
\\&\hspace*{-.4cm}
   + 2 \lambda_4\ \big(\chi_L^\dag \chi_L\big)\ \big(\chi_R^\dag\chi_R\big)
   + 2 \alpha_1 {\rm Tr} \big[\phi^\dag \phi\big] \big[\chi_L^\dag \chi_L \!+\! \chi_R^\dag \chi_R\big]
\\&\hspace*{-.4cm}
   + 2 \alpha_2 \big[ \big(\chi_L^\dag \phi\big) \big(\chi_L\phi^\dag\big) +
          \big(\phi^\dag \chi_R^\dag\big)\ \big(\phi\chi_R\big)\big]
\\&\hspace*{-.4cm}
+ 2 \alpha_3 \big[ \big(\chi_L^\dag \hat\phi^\dag\big)\
          \big(\chi_L\hat\phi\big) + \big(\hat\phi\chi_R^\dag\big)\
          \big(\hat\phi^\dag\chi_R\big)\big]\,.
\esp\ee
Imposing consistent vacuum stability conditions and non-tachyonic states, phenomenologically acceptable scenarios always feature $\alpha\equiv\alpha_1\simeq\alpha_2=\alpha_3$ and $\lambda_2=0$.

The breaking of the left-right symmetry down to electromagnetism yields a vacuum configuration in which all
neutral components of the scalar fields acquire non-vanishing vacuum expectation values (vevs), with the exception of $\phi^0_1$, which is protected by the conservation of the generalised lepton number $L$ (which also forbids mixing between the SM $d$ and exotic $d^\prime$ quarks),
\be
  \langle \phi  \rangle = \bpm 0&0\\0 & \frac{k}{\sqrt{2}} \epm \!, ~~
  \langle \chi_L\rangle = \bpm 0\\ \frac{\vL}{\sqrt{2}} \epm \!,~~
  \langle \chi_R\rangle = \bpm 0\\ \frac{\vR}{\sqrt{2}} \epm \,.
\ee
Subsequently, while the (pseudo)scalar components of $\phi^0_1$ is disallowed from further mixing with any other (pseudo)scalar degree of freedom, the rest of the neutral Higgs bosons do. We refer to \cite{Frank:2019nid} for expressions of the corresponding $3\times 3$ mixing matrices that are entirely determined by the other parameters of the model (including those introduced below). In the charged sector, the physical massive charged Higgs bosons $H_1^\pm$ and $H_2^\pm$, and the two massless Goldstone bosons $G_1^\pm$ and $G_2^\pm$ to be absorbed by the $W\!\equiv\! W_L$ and $W_R$ charged gauge bosons, are defined by
\be\bsp
 \bpm \phi_2^\pm\\\chi_L^\pm\epm \!=\!
 \bpm \cos\beta & \sin\beta\\ -\sin\beta & \cos\beta\epm
 \bpm H_1^\pm\\G_1^\pm\epm \ \ \text{with}\ \  \tan\beta \!=\! \frac{k}{\vL}\,,\\[.1cm]
 \bpm \phi_1^\pm\\\chi_R^\pm\epm \!=\!
 \bpm \cos\zeta & \sin\zeta\\ -\sin\zeta & \cos\zeta\epm
 \bpm H_2^\pm\\G_2^\pm\epm  \ \ \text{with}\ \  \tan\zeta \!=\! \frac{k}{\vR}\,,
\esp\ee
Correspondingly, the squared charged scalar masses, that are relevant for loop-induced FCNC top quark decays, read
\be
  M_{H_1^\pm}^2 = -\frac{\kappa \vR (k^2+\vL^2)}{\sqrt{2} k \vL}\,,\quad \
  M_{H_2^\pm}^2 = -\frac{\kappa \vL (k^2+\vR^2)}{\sqrt{2} k \vR}\,.
\label {eq:mh2squar}\ee
As $v^2\equiv k^2+v_L^2 \ge 2kv_L$ and $v^{\prime \,2} \equiv k^2+v_R^2 \simeq v_R^2$, the $H_1^\pm$ state is generally heavy, while the $H_2^\pm$ one lies in the sub-TeV to TeV range for a large part of the parameter space.

The breaking of the left-right symmetry generates masses for the gauge bosons and induces their mixing. Due to the conservation of the generalised lepton number (preventing $\phi_1^0$ from acquiring a vev), the charged $W$ and $W_R$ bosons do not mix. In contrast, the neutral bosons undergo mixing, which results in a massless photon and two massive $Z$ and $Z'$ states. The masses of the extra gauge bosons are 
\be
  M_{W_R} \!=\! \frac12 \gR v' \,,\ \
  M_{Z'} \!=\! \frac12 \sqrt{\gBL^2 s_{\phw}^2 \vL^2 \!+\! \frac{\gR^2 (c_{\phw}^4 k^2 \!+\! \vR^2)}{c_{\phw}^2}} \,.
\ee
These expressions depend on the sine ($s_i$) and cosine ($c_i$) of the two mixing angles, that are determined from
\be
  s_{\phw} = \frac{\gBL}{\sqrt{\gBL^2+\gR^2}} = \frac{\gY}{\gR}\,, \quad
  s_{\tw}  = \frac{\gY}{\sqrt{\gL^2+\gY^2}} = \frac{e}{\gL} \ ,
  \label{eq:couplings}
\ee
where $\gY$ and $e$ denote the hypercharge and electromagnetic coupling constant respectively. The mass ratio of the two $SU(2)_{R'}$(-like) gauge bosons satisfies $M_{W_R}/M_{Z^\prime} \sim c_{\phw}$. Thus for values of $g_R \ne g_L$, this ratio can be substantially large or small. This is the same as in the LRSM, however there are independent collider limits on $M_{W_R}$ from decays into quarks or lepton-neutrino which do not apply here. Thus the $W_R$ boson can be much lighter than the $Z^\prime$.

%%%%%%%%%%%%%%%%%%%%%%%%%%%%%%%%%%%%%%%%
\section{Rare top decays in the ALRM}
\label{sec:topraredecay}
%%%%%%%%%%%%%%%%%%%%%%%%%%%%%%%%%%%%%%%%
In the SM, FCNC couplings arise at one loop from exchanges of $W$ bosons and down-type quarks. The same mechanism applies to ALRM scenarios, with additional contributions from exchanges of charged Higgs bosons $H_{1,2}^\pm$, charged $SU(2)_{R'}$ bosons $W_R$, and $d^\prime$-type quarks. Among these contributions, loops featuring exotic quarks ($d^\prime$, $s^\prime$ and $b^\prime$) and $W_R$ and $H_2^\pm$ bosons dominate over loops involving ordinary quarks ($d$, $s$ and $b$) and $W$ and $H_1^\pm$ states. This dominance originates from the fact that right-handed (left-handed) top quarks directly couple to exotic $d_R^\prime$ quarks and not to ordinary $d_R$-type quarks through interactions with the $W_R$ ($H_2^\pm$) boson, from larger relevant Yukawa couplings, and from the absence of any GIM suppression in the BSM sector.

%%%%%%%%%%%%%%%%%%%%
\begin{table}
  \centering
  \setlength\tabcolsep{6pt}\renewcommand{\arraystretch}{1.1}
  \resizebox{\columnwidth}{!}{
    \begin{tabular}{cc|cc}
    Parameter  & Range & Parameter & Range \\
    \hline
    $\tan\beta$ & $[0.7, 50]$       & $\alpha$    & $[ 0.01, 0.5]$ \\
    $\gR$       & $[0.37, 0.768]$   & $\lambda_3$ & $[0.01, 0.09]$\\    
    $v'$ [TeV]  & $[6.5, 20]$       & $\kappa$ [GeV]   & $[-1000, -1]$\\
    $\theta_{12}^{'}$, $\theta_{13}^{'}$, $\theta_{23}^{'}$ & $[0, \pi/4]$ 
      & $m_{d'}$, $m_{s'}$, $m_{b'}$ [GeV] & $[100, 10000]$\\
  \end{tabular}}
  \caption{Set of independent parameters driving our ALRM exploration, shown with the range of numerical values used in our scan.}
  \label{tab:scan_lim}
\end{table}
%%%%%%%%%%%%%%%%%%%%

In order to explore the model's parameter space, we consider the independent parameters relevant for rare top decays listed in Table~\ref{tab:scan_lim}, and in terms of which all other Lagrangian parameters and physical masses or mixings can be expressed. These parameters include the $SU(2)_{R'}$ gauge coupling $\gR$ and CKM mixing matrix that we express in terms of inter-generational mixing angles $\theta_{12}'$, $\theta_{13}'$ and $\theta_{23}'$ (ignoring any possible $CP$ violating phase as irrelevant for rare top decays); the vev $v'$ and mixing angle $\tan\beta$; the scalar potential parameters $\lambda_3$, $\kappa$ and $\alpha$; and the masses of the exotic quarks $m_{d'}$, $m_{s'}$ and $~m_{b'}$.

To compute the branching ratios for all possible top FCNC decays, we rely on \texttt{FeynArts}~\cite{Hahn:2010zi} and \texttt{FormCalc}~\cite{Hahn:2016ebn} for the generation of Feynman diagrams, the derivation of associated matrix elements, and their conversion into a Fortran package for numerical evaluation. The ALRM model implementation in \texttt{FeynArts} has been performed with \texttt{FeynRules}~\cite{Christensen:2009jx, Alloul:2013bka} using the model file developed in~\cite{Frank:2019nid}.  As a cross-check, we verified that SM predictions were retrieved after decoupling the BSM sector. Our numerical analysis was conducted by integrating the Fortran output of \texttt{FormCalc} into a custom scan module that explores the parameter space defined in Table~\ref{tab:scan_lim}, within the specified ranges. 

Our selection criteria for the parameter space regions to be scanned aims to ensure that at least one of the charged Higgs bosons, the extra gauge bosons, and the exotic quarks all fall within the sub-TeV to TeV range. Additionally, we allow the mixing angles $\theta_{ij}'$, which parameterise exotic quark mixings, to vary freely from 0 (no-mixing) to $\pi/4$ (maximum mixing). Given the similarity in masses of exotic quarks, our results are, however, not expected to be significantly dependent on $\theta_{ij}'$. The $SU(2)_{R'}$ coupling $\gR$ is allowed to vary within the range $[0.37, 0.768]$, where the lower bound ensures that $\gR/\gL$ exceeds $\tan\tw$~\cite{Dev:2016dja} and the upper bound (approximately $\sqrt{4\pi}$) maintains $\gR$ within the perturbative regime.\footnote{ 
 The minimum value of $g_R$ is easily obtained from Eq.~\eqref{eq:couplings}, that yields $\sin \phi =\frac {g_L \tan \theta}{g_R}$. Requiring $\sin \phi \le 1$ gives $g_R/g_L \ge \tan \theta$, thus the minimum value of $g_R=0.37$~\cite{Frank:2020odd}.}

In addition, the SM parameters are fixed according to the Particle Data Group Review~\cite{ParticleDataGroup:2022pth}, except for the total decay width of the top quark, which is set to $\Gamma_t=1.32 \pm 0.5$ GeV as in~\cite{ATLAS:2019onj}. In the scan, we randomly sample $10^{7}$ scenarios, ensuring that $M_{Z^{\prime}} \geq 4.5$~TeV to comply with LHC constraints~\cite{CMS:2018ipm,ATLAS:2017fih}, and that $M_{H_2^\pm}\geq 90$~GeV to satisfy LEP limits~\cite{OPAL:1998znn}. Furthermore, scenarios with $M_{H_2^\pm} > 100$~GeV are subjected to an additional condition, requiring that the associated production cross section at the LHC is less than 0.1~pb, aligning with existing limits. 

{\renewcommand{\arraystretch}{1.2}  
\begin{table}\setlength\tabcolsep{10pt}
\begin{tabular}{ ccc }
 Channel & SM & Max. BR in ALRM\\
 \hline
 $t\rightarrow c \gamma$ & $1.0 \times 10^{-13}$  & $3.4\times 10^{-11}$  \\
 $t\rightarrow u\gamma$ & $8.1 \times 10^{-16}$  &   $7.1\times 10^{-12}$ \\
 $t\rightarrow c Z$ & $2.4\times 10^{-14}$ & $1.1\times 10^{-10}$  \\
 $t\rightarrow u Z$ & $1.9\times 10^{-16}$ & $7.9\times 10^{-11}$\\
 $t\rightarrow c H$ & $1.8 \times 10^{-13}$ & $7.8\times 10^{-9}$  \\
 $t\rightarrow u H$ & $1.4\times 10^{-15}$  & $8.3\times 10^{-9}$ \\
 $t\rightarrow c g$ & $8.5 \times 10^{-12}$ & $2.9\times 10^{-10}$ \\
 $t\rightarrow u g$ & $6.6 \times 10^{-14}$  & $4.3\times 10^{-11}$ \\
\end{tabular}
\caption{Maximum branching ratios obtained in the scan of the ALRM parameter space defined in Table~\ref{tab:scan_lim}. SM predictions for each channel are given for comparison~\cite{ParticleDataGroup:2022pth}.} \label{tab:maxBR}
\end{table}
}

\begin{figure}
  \includegraphics[width=\columnwidth]{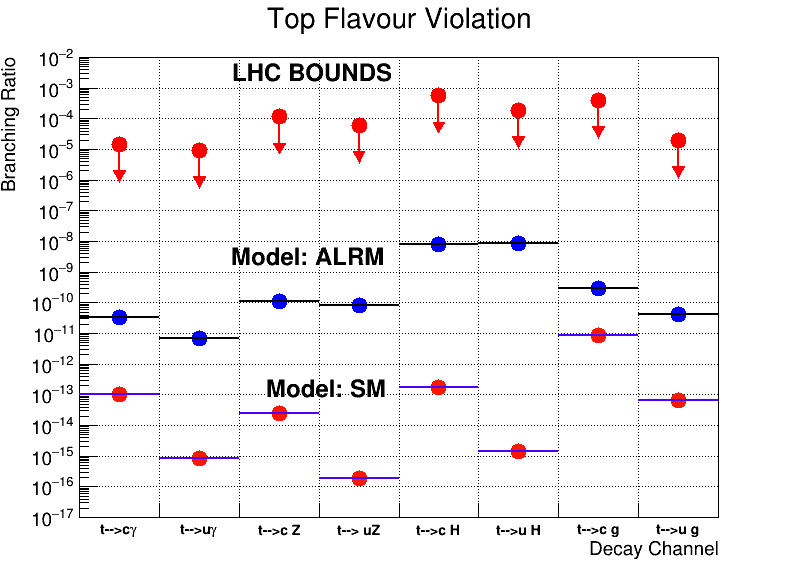}
  \caption{Maximum branching ratios for the considered rare top decays obtained from our ALRM scan (blue dots) and in the SM~\cite{ParticleDataGroup:2022pth} (red dots). For comparison, we include current bounds from LHC data (red dots with arrows)~\cite{CMS:2016uzc, CMS:2017wcz, CMS:2017twu, CMS:2021hug, CMS:2023tir}.}
  \label{fig:cmsplot}
\end{figure}

The objective of our scan is to identify regions in the parameter space that maximise individual branching ratios associated with the rare top decays $t \!\to\! q \gamma$, $q g$, $q Z$ and $q H$ (with $q\!=\!u, c$). The largest obtained branching ratios are collected in Table~\ref{tab:maxBR}, along with the corresponding SM predictions. We observe that branching ratios can reach up to $10^{-11} - 10^{-10}$ for decays into spin-1 bosons and approach $10^{-8}$ for decays into a Higgs boson. These results are also visually presented in Fig.~\ref{fig:cmsplot}, where we include the current experimental limits~\cite{CMS:2016uzc, CMS:2017wcz, CMS:2017twu, CMS:2021hug, CMS:2023tir}. These experimental bounds can be further compared with the expected improvement of about one order of magnitude from the high-luminosity run of the LHC, and the additional order of magnitude anticipated from a future 100~TeV hadronic collider~\cite{ATLAS:2016qxw, Cremer:2023gne}.  

\begin{figure*}
\centering
  \includegraphics[width=0.325\textwidth,trim={2 3 30 5},clip]{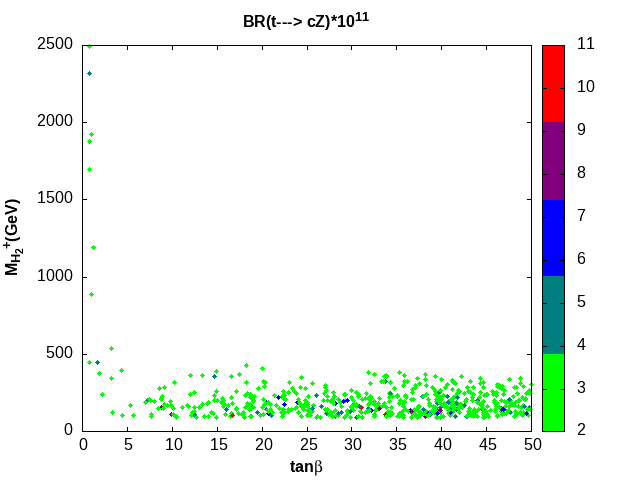}
  \includegraphics[width=0.325\textwidth,trim={2 3 30 5},clip]{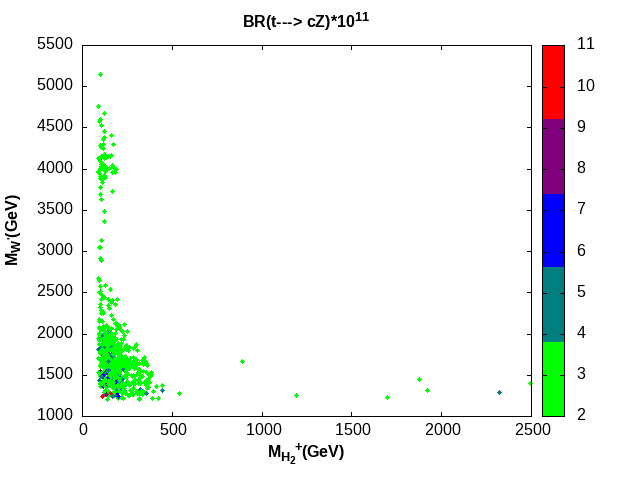}
  \includegraphics[width=0.325\textwidth,trim={2 3 30 5},clip]{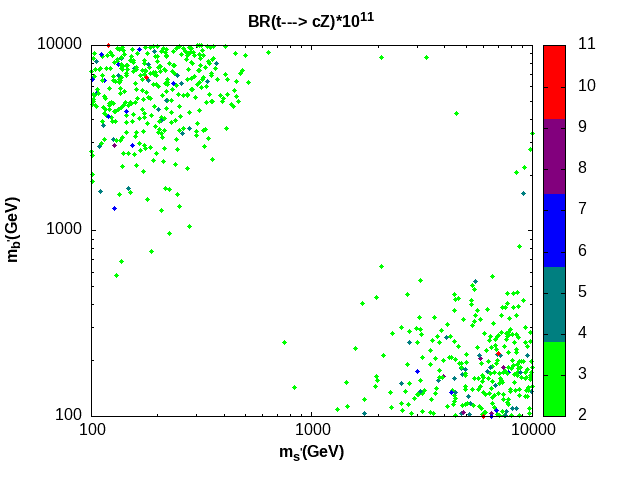}\\
  \includegraphics[width=0.325\textwidth,trim={2 3 30 5},clip]{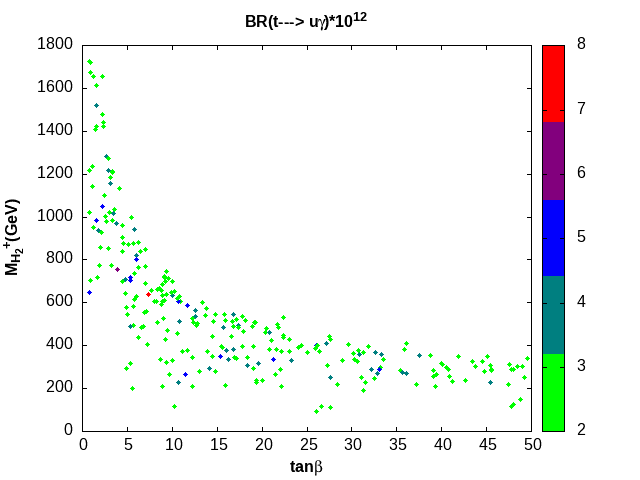}
  \includegraphics[width=0.325\textwidth,trim={2 3 30 5},clip]{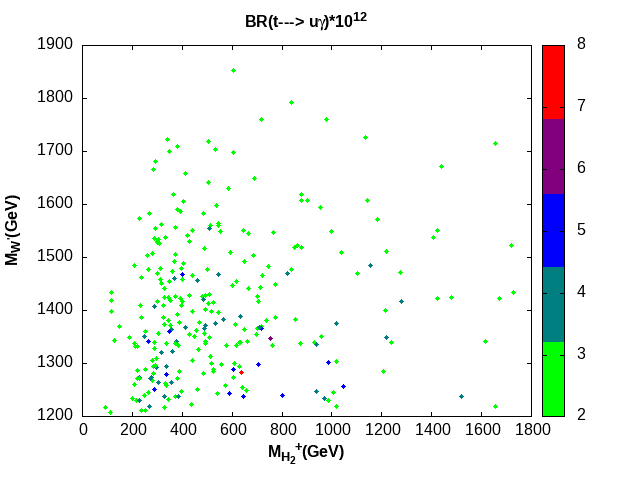}
  \includegraphics[width=0.325\textwidth,trim={2 3 30 5},clip]{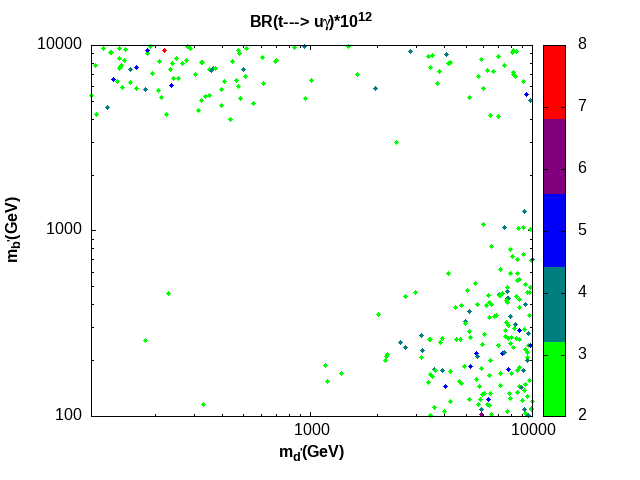}
\caption{\label{fig:mh2mw}Correlations between various pairs of ALRM parameters, as returned by our scan, for scenarios maximising branching ratios associated with $t\to cZ$ (top row) and $t\to u\gamma$ (bottom row) decays. We project our results in the $(\tan\beta, M_{H_2^\pm})$ plane (left), $(M_{H_2^\pm}, M_{W_R})$ plane (middle), and $(m_{d'}, m_{b'})$ plane (right), after the removal of scenarios yielding too small branching ratios.}
\end{figure*}

These results unveil that the decays $t \to q \gamma$ and $t \to g \gamma$ for $q=u, c$ exhibit a pronounced dependence on the final quark flavour, closely intertwined with the mass values of the additional states.  In the case of decays into up quarks, that is illustrated with the plots in the bottom row of Fig.~\ref{fig:mh2mw}, significant branching ratios favour $W_R$-boson masses in the 1.2 -- 1.5 TeV range, correlating with a charged Higgs boson of mass ranging from a few hundred GeV to 1.5~TeV. Such a light $W_R$ boson additionally influences the right-handed CKM mixing angle $\theta_{13}^\prime$, requiring it to be at least $25^\circ$ and below $45^\circ$ (which corresponds to a maximally mixed situation), and pushes exotic quark masses deep into the multi-TeV regime. Furthermore, a TeV-scale charged Higgs boson constrains the possible values of $\tan\beta$ as driven by Eq.~\eqref{eq:mh2squar}. Notably, charged Higgs boson masses beyond 1~TeV imply $\tan\beta \lesssim 5$, while a lighter charged Higgs boson below 1~TeV allows for substantial values for ${\cal B}(t \!\to\! u \gamma)$ and ${\cal B}(t \to u g)$ only with a large $\tan\beta$ value. On the other hand, large enhancements of ${\cal B}(t \!\to\! c \gamma)$ and ${\cal B}(t \!\to\! c g)$ favour $W_R$-boson masses in the 3.5 -- 4 TeV regime, which is compensated by a charged Higgs boson mass between 90 and 115~GeV. Due to the large $W_R$ mass, these enhancements are independent of the right-handed CKM mixing angles and exotic quark masses, while a light charged Higgs boson remains uncorrelated with any constraint on $\tan\beta$.

Predictions for top decays into $uZ$ and $cZ$ final states exhibit very similar properties, as illustrated for $t \!\to\! cZ$ in the plots shown in the top row of Figure~\ref{fig:mh2mw}. The $W_R$ mass needs to be either within 1.2 -- 2~TeV or around 4~TeV. In the former case, the charged Higgs boson must have a mass greater than 500~GeV, and $\tan\beta\lesssim 5$, while in the latter case, $M_{H_2^\pm}$ has to be between 90 and 115~GeV, with $\tan\beta$ remaining unconstrained. For decays into a first (second) generation up-type quark, the right-handed CKM mixing angle $\theta_{13}^\prime$ ($\theta_{23}^\prime$) is required to be in the $25^\circ - 45^\circ$ range, with the other two angles allowed to vary freely, up to unitarity constraints.

The decay processes $t\!\to\! uH$ and $t\!\to\! cH$ have properties quite similar to those involving a final state with a $Z$ boson. The main difference is that the $W_R$ boson has to be light, with a mass below 2~TeV, to guarantee an enhanced branching ratios ${\cal B}(t \!\to\! q H)$, and the charged Higgs boson $H_2^\pm$ must be heavy with a mass of a few hundred GeV. Furthermore, $\tan\beta$ is unrestricted, although it has to be smaller than 10 when the charged Higgs mass is larger than 500~GeV.

Finally, it is important to note that, in all cases, the mass of the $Z'$ boson must be consistent with experimental data. Therefore, it satisfies $M_{Z^\prime} \approx 4.5-5$~TeV for all classes of decays studied.

The relative masses of the charged Higgs boson $H_2^\pm$ and of the gauge boson $W_R$ determine the significance of the different diagrams contributing to the studied rare top decays. When $M_{H_2^\pm}$ is in the 100~GeV range, diagrams involving a charged-Higgs exchange dominate. In this case, the distinction between decays into first-generation and second-generation quarks is linked to the SM quark mass factor that multiplies specific terms in the corresponding amplitudes. Consequently, charged Higgs-boson exchanges play a more significant role in decays involving charm quarks as $m_c \gg m_u$. While other terms in these amplitudes are multiplied by a factor of the mass of the exotic quarks exchanged, all these masses are in the same ballpark, thus introducing no substantial difference between decays into first and second-generation quarks. Scenarios yielding an enhanced branching ratio for one of the rare top decays considered indeed always feature one lower exotic quark mass of about 1~TeV and two larger masses of about 10~TeV.

%%%%%%%%%%%%%%%%%%%%%%%%%%%%%%%%%%%%%%%% 
\section{ Conclusions}
\label{sec:conclusion}
%%%%%%%%%%%%%%%%%%%%%%%%%%%%%%%%%%%%%%%% 

Rare top decays involving FCNCs are highly suppressed in the SM due to the GIM mechanism, with branching ratios of $t \rightarrow q X$ (with $q = u, c$ and $X = \gamma, g, Z, H$) of the order of $10^{-12} - 10^{-16}$. A significant enhancement in these rates would, therefore, be a signal of new physics. In this work, we examined rare FCNC top quark decays in the ALRM. In addition to extra $SU(2)_{R'}$ gauge bosons and exotic quarks, this model also contains new charged Higgs bosons. Unlike in the usual left-right model, these charged Higgs states do not lead to FCNCs at tree level and are thus allowed to be light. We analysed the effect of all model's parameters on the corresponding branching ratios, and we explored the consequences of large deviations on the parameter space. We found that an enhancement of $10^2 - 10^6$ compared to the SM is achievable while imposing consistency with current theoretical and experimental constraints.  

Notably, we observed an enhancement of $10^6$ for the branching ratios $\mathcal{B}(t\!\to\! uZ)$ and $\mathcal{B}(t\!\to\! uH)$, while for other decay processes, the enhancement lies in the range of $10^{2} - 10^{4}$. Two significant classes of scenarios lead to such a large enhancement. The first features a charged Higgs boson $H_2^\pm$ with a mass greater than 500~GeV together with $\tan\beta\lesssim 5$. The second exhibits a light charged Higgs boson with a mass of about 100~GeV, which is possible if $\tan\beta>5$. Related to these findings, the charged gauge boson $W_R$ can possibly be light, with a mass  in the 1.2 -- 2~TeV regime, or heavy, with a mass of about 4~TeV. Additionally, one of the right-handed CKM mixing angles has to be large ($\theta^\prime_{23}$ for decays into charm quarks and $\theta^\prime_{13}$ for decays into up quarks), with the other angles generally small as required by the unitarity of the right-handed CKM matrix. While these enhancements are significant, they fall below observability prospects of a future hadronic collider expected to operate at a centre-of-mass energy of 100~TeV.

Studying rare top decays therefore provides an interesting avenue to explore the ALRM. It serves as indirect probes for the possibility of a light charged Higgs boson or moderately heavy $W_R$ boson (both currently allowed by data), as well as TeV-scale exotic quarks. It will also allow for an unambiguous distinction of the ALRM from usual LRSM setups in which such a mass spectrum is excluded.

\section*{Acknowledgements}  
The work of MF has been partly supported by NSERC through grant number SAP105354, and that of BF by the French \emph{Agence Nationale de la Recherche} (ANR) through grant ANR-21-CE31-0013 (DMwithLLPatLHC). The work of SKG and PP has been supported by SERB, DST, India through grants TAR/2023/000116 and CRG/2022/002670 respectively. We thank Thomas Hahn for providing support while setting up our numerical code relying on \texttt{FeynArts} and \texttt{FormCalc}.

\bibliographystyle{JHEP}
\bibliography{FCNC_ALRM_PLB}

\providecommand{\href}[2]{#2}\begingroup\raggedright\begin{thebibliography}{10}

\bibitem{D0:1995jca}
{\scshape D0} collaboration, S.~Abachi et~al., \emph{{Observation of the top
  quark}}, \href{http://dx.doi.org/10.1103/PhysRevLett.74.2632}{\emph{Phys.
  Rev. Lett.} {\bf 74} (1995) 2632--2637},
  [\href{http://arxiv.org/abs/hep-ex/9503003}{{\tt hep-ex/9503003}}].

\bibitem{CDF:1995wbb}
{\scshape CDF} collaboration, F.~Abe et~al., \emph{{Observation of top quark
  production in $\bar{p}p$ collisions}},
  \href{http://dx.doi.org/10.1103/PhysRevLett.74.2626}{\emph{Phys. Rev. Lett.}
  {\bf 74} (1995) 2626--2631}, [\href{http://arxiv.org/abs/hep-ex/9503002}{{\tt
  hep-ex/9503002}}].

\bibitem{TopQuarkWorkingGroup:2013hxj}
{\scshape Top Quark Working Group} collaboration, K.~Agashe et~al.,
  \emph{{Working Group Report: Top Quark}},  in \emph{{Snowmass 2013}:
  {Snowmass on the Mississippi}}, 11, 2013.
\newblock \href{http://arxiv.org/abs/1311.2028}{{\tt 1311.2028}}.

\bibitem{Mele:1998ag}
B.~Mele, S.~Petrarca and A.~Soddu, \emph{{A New evaluation of the t
  ---\ensuremath{>} cH decay width in the standard model}},
  \href{http://dx.doi.org/10.1016/S0370-2693(98)00822-3}{\emph{Phys. Lett. B}
  {\bf 435} (1998) 401--406}, [\href{http://arxiv.org/abs/hep-ph/9805498}{{\tt
  hep-ph/9805498}}].

\bibitem{Aguilar-Saavedra:2002lwv}
J.~A. Aguilar-Saavedra and B.~M. Nobre, \emph{{Rare top decays t
  ---\ensuremath{>} c gamma, t ---\ensuremath{>} cg and CKM unitarity}},
  \href{http://dx.doi.org/10.1016/S0370-2693(02)03230-6}{\emph{Phys. Lett. B}
  {\bf 553} (2003) 251--260}, [\href{http://arxiv.org/abs/hep-ph/0210360}{{\tt
  hep-ph/0210360}}].

\bibitem{Aguilar-Saavedra:2004mfd}
J.~A. Aguilar-Saavedra, \emph{{Top flavor-changing neutral interactions:
  Theoretical expectations and experimental detection}}, {\emph{Acta Phys.
  Polon. B} {\bf 35} (2004) 2695--2710},
  [\href{http://arxiv.org/abs/hep-ph/0409342}{{\tt hep-ph/0409342}}].

\bibitem{Zhang:2013xya}
C.~Zhang and F.~Maltoni, \emph{{Top-quark decay into Higgs boson and a light
  quark at next-to-leading order in QCD}},
  \href{http://dx.doi.org/10.1103/PhysRevD.88.054005}{\emph{Phys. Rev. D} {\bf
  88} (2013) 054005}, [\href{http://arxiv.org/abs/1305.7386}{{\tt 1305.7386}}].

\bibitem{Durieux:2014xla}
G.~Durieux, F.~Maltoni and C.~Zhang, \emph{{Global approach to top-quark
  flavor-changing interactions}},
  \href{http://dx.doi.org/10.1103/PhysRevD.91.074017}{\emph{Phys. Rev. D} {\bf
  91} (2015) 074017}, [\href{http://arxiv.org/abs/1412.7166}{{\tt 1412.7166}}].

\bibitem{Castro:2022qkg}
N.~F. Castro and K.~Skovpen, \emph{{Flavour-Changing Neutral Scalar
  Interactions of the Top Quark}},
  \href{http://dx.doi.org/10.3390/universe8110609}{\emph{Universe} {\bf 8}
  (2022) 609}, [\href{http://arxiv.org/abs/2210.09641}{{\tt 2210.09641}}].

\bibitem{Larios:2006pb}
F.~Larios, R.~Martinez and M.~A. Perez, \emph{{New physics effects in the
  flavor-changing neutral couplings of the top quark}},
  \href{http://dx.doi.org/10.1142/S0217751X06033039}{\emph{Int. J. Mod. Phys.
  A} {\bf 21} (2006) 3473--3494},
  [\href{http://arxiv.org/abs/hep-ph/0605003}{{\tt hep-ph/0605003}}].

\bibitem{Barros:2019wxe}
M.~Barros, N.~F. Castro, J.~Erdmann, G.~Ge\ss{}ner, K.~Kr\"oninger,
  S.~La~Cagnina et~al., \emph{{Study of interference effects in the search for
  flavour-changing neutral current interactions involving the top quark and a
  photon or a Z boson at the LHC}},
  \href{http://dx.doi.org/10.1140/epjp/s13360-020-00346-3}{\emph{Eur. Phys. J.
  Plus} {\bf 135} (2020) 339}, [\href{http://arxiv.org/abs/1909.08443}{{\tt
  1909.08443}}].

\bibitem{Diaz-Cruz:1989tem}
J.~L. Diaz-Cruz, R.~Martinez, M.~A. Perez and A.~Rosado, \emph{{Flavor Changing
  Radiative Decay of The T Quark}},
  \href{http://dx.doi.org/10.1103/PhysRevD.41.891}{\emph{Phys. Rev. D} {\bf 41}
  (1990) 891--894}.

\bibitem{Eilam:1990zc}
G.~Eilam, J.~L. Hewett and A.~Soni, \emph{{Rare decays of the top quark in the
  standard and two Higgs doublet models}},
  \href{http://dx.doi.org/10.1103/PhysRevD.44.1473}{\emph{Phys. Rev. D} {\bf
  44} (1991) 1473--1484}.

\bibitem{Atwood:1996vj}
D.~Atwood, L.~Reina and A.~Soni, \emph{{Phenomenology of two Higgs doublet
  models with flavor changing neutral currents}},
  \href{http://dx.doi.org/10.1103/PhysRevD.55.3156}{\emph{Phys. Rev. D} {\bf
  55} (1997) 3156--3176}, [\href{http://arxiv.org/abs/hep-ph/9609279}{{\tt
  hep-ph/9609279}}].

\bibitem{Botella:2015hoa}
F.~J. Botella, G.~C. Branco, M.~Nebot and M.~N. Rebelo, \emph{{Flavour Changing
  Higgs Couplings in a Class of Two Higgs Doublet Models}},
  \href{http://dx.doi.org/10.1140/epjc/s10052-016-3993-0}{\emph{Eur. Phys. J.
  C} {\bf 76} (2016) 161}, [\href{http://arxiv.org/abs/1508.05101}{{\tt
  1508.05101}}].

\bibitem{Abbas:2015cua}
G.~Abbas, A.~Celis, X.-Q. Li, J.~Lu and A.~Pich, \emph{{Flavour-changing top
  decays in the aligned two-Higgs-doublet model}},
  \href{http://dx.doi.org/10.1007/JHEP06(2015)005}{\emph{JHEP} {\bf 06} (2015)
  005}, [\href{http://arxiv.org/abs/1503.06423}{{\tt 1503.06423}}].

\bibitem{Baum:2008qm}
I.~Baum, G.~Eilam and S.~Bar-Shalom, \emph{{Scalar flavor changing neutral
  currents and rare top quark decays in a two H iggs doublet model 'for the top
  quark'}}, \href{http://dx.doi.org/10.1103/PhysRevD.77.113008}{\emph{Phys.
  Rev. D} {\bf 77} (2008) 113008}, [\href{http://arxiv.org/abs/0802.2622}{{\tt
  0802.2622}}].

\bibitem{Dedes:2014asa}
A.~Dedes, M.~Paraskevas, J.~Rosiek, K.~Suxho and K.~Tamvakis, \emph{{Rare
  Top-quark Decays to Higgs boson in MSSM}},
  \href{http://dx.doi.org/10.1007/JHEP11(2014)137}{\emph{JHEP} {\bf 11} (2014)
  137}, [\href{http://arxiv.org/abs/1409.6546}{{\tt 1409.6546}}].

\bibitem{Cao:2007dk}
J.~J. Cao, G.~Eilam, M.~Frank, K.~Hikasa, G.~L. Liu, I.~Turan et~al.,
  \emph{{SUSY-induced FCNC top-quark processes at the large hadron collider}},
  \href{http://dx.doi.org/10.1103/PhysRevD.75.075021}{\emph{Phys. Rev. D} {\bf
  75} (2007) 075021}, [\href{http://arxiv.org/abs/hep-ph/0702264}{{\tt
  hep-ph/0702264}}].

\bibitem{Lopez:1997xv}
J.~L. Lopez, D.~V. Nanopoulos and R.~Rangarajan, \emph{{New supersymmetric
  contributions to $t \to c$ V}},
  \href{http://dx.doi.org/10.1103/PhysRevD.56.3100}{\emph{Phys. Rev. D} {\bf
  56} (1997) 3100--3106}, [\href{http://arxiv.org/abs/hep-ph/9702350}{{\tt
  hep-ph/9702350}}].

\bibitem{Eilam:2001dh}
G.~Eilam, A.~Gemintern, T.~Han, J.~M. Yang and X.~Zhang, \emph{{Top quark rare
  decay t ---\ensuremath{>} ch in R-parity violating SUSY}},
  \href{http://dx.doi.org/10.1016/S0370-2693(01)00598-6}{\emph{Phys. Lett. B}
  {\bf 510} (2001) 227--235}, [\href{http://arxiv.org/abs/hep-ph/0102037}{{\tt
  hep-ph/0102037}}].

\bibitem{Gaitan:2004by}
R.~Gaitan, O.~G. Miranda and L.~G. Cabral-Rosetti, \emph{{Rare top and Higgs
  decays in alternative left-right symmetric models}},
  \href{http://dx.doi.org/10.1103/PhysRevD.72.034018}{\emph{Phys. Rev. D} {\bf
  72} (2005) 034018}, [\href{http://arxiv.org/abs/hep-ph/0410268}{{\tt
  hep-ph/0410268}}].

\bibitem{Frank:2005vd}
M.~Frank and I.~Turan, \emph{{$t \to$ cg, $c \gamma$, cZ in the left-right
  supersymmetric model}},
  \href{http://dx.doi.org/10.1103/PhysRevD.72.035008}{\emph{Phys. Rev. D} {\bf
  72} (2005) 035008}, [\href{http://arxiv.org/abs/hep-ph/0506197}{{\tt
  hep-ph/0506197}}].

\bibitem{Dey:2016cve}
U.~K. Dey and T.~Jha, \emph{{Rare top decays in minimal and nonminimal
  universal extra dimension models}},
  \href{http://dx.doi.org/10.1103/PhysRevD.94.056011}{\emph{Phys. Rev. D} {\bf
  94} (2016) 056011}, [\href{http://arxiv.org/abs/1602.03286}{{\tt
  1602.03286}}].

\bibitem{Gao:2013fxa}
T.-J. Gao, T.-F. Feng and J.-B. Chen, \emph{{$t \to c \gamma$ and $t \to c g$
  in warped extra dimensions}},
  \href{http://dx.doi.org/10.1007/JHEP02(2013)029}{\emph{JHEP} {\bf 02} (2013)
  029}, [\href{http://arxiv.org/abs/1303.0082}{{\tt 1303.0082}}].

\bibitem{Diaz-Furlong:2016ril}
A.~Diaz-Furlong, M.~Frank, N.~Pourtolami, M.~Toharia and R.~Xoxocotzi,
  \emph{{Flavor-changing decays of the top quark in 5D warped models}},
  \href{http://dx.doi.org/10.1103/PhysRevD.94.036001}{\emph{Phys. Rev. D} {\bf
  94} (2016) 036001}, [\href{http://arxiv.org/abs/1603.08929}{{\tt
  1603.08929}}].

\bibitem{Liu:2021crr}
Y.~Liu, B.~Yan and R.~Zhang, \emph{{Loop induced top quark FCNC through top
  quark and dark matter interactions}},
  \href{http://dx.doi.org/10.1016/j.physletb.2022.136964}{\emph{Phys. Lett. B}
  {\bf 827} (2022) 136964}, [\href{http://arxiv.org/abs/2103.07859}{{\tt
  2103.07859}}].

\bibitem{Crivellin:2022fdf}
A.~Crivellin, M.~Kirk, T.~Kitahara and F.~Mescia, \emph{{Large
  t\textrightarrow{}cZ as a sign of vectorlike quarks in light of the W mass}},
  \href{http://dx.doi.org/10.1103/PhysRevD.106.L031704}{\emph{Phys. Rev. D}
  {\bf 106} (2022) L031704}, [\href{http://arxiv.org/abs/2204.05962}{{\tt
  2204.05962}}].

\bibitem{Banerjee:2018fsx}
S.~Banerjee, M.~Chala and M.~Spannowsky, \emph{{Top quark FCNCs in extended
  Higgs sectors}},
  \href{http://dx.doi.org/10.1140/epjc/s10052-018-6150-0}{\emph{Eur. Phys. J.
  C} {\bf 78} (2018) 683}, [\href{http://arxiv.org/abs/1806.02836}{{\tt
  1806.02836}}].

\bibitem{Aguilar-Saavedra:2008nuh}
J.~A. Aguilar-Saavedra, \emph{{A Minimal set of top anomalous couplings}},
  \href{http://dx.doi.org/10.1016/j.nuclphysb.2008.12.012}{\emph{Nucl. Phys. B}
  {\bf 812} (2009) 181--204}, [\href{http://arxiv.org/abs/0811.3842}{{\tt
  0811.3842}}].

\bibitem{Agram:2013koa}
J.-L. Agram, J.~Andrea, E.~Conte, B.~Fuks, D.~Gel\'e and P.~Lansonneur,
  \emph{{Probing top anomalous couplings at the LHC with trilepton signatures
  in the single top mode}},
  \href{http://dx.doi.org/10.1016/j.physletb.2013.06.052}{\emph{Phys. Lett. B}
  {\bf 725} (2013) 123--126}, [\href{http://arxiv.org/abs/1304.5551}{{\tt
  1304.5551}}].

\bibitem{Oyulmaz:2018irs}
K.~Y. Oyulmaz, A.~Senol, H.~Denizli, A.~Yilmaz, I.~Turk~Cakir and O.~Cakir,
  \emph{{Probing anomalous $tq\gamma$ and $tqg$ couplings via single top
  production in association with photon at FCC-hh}},
  \href{http://dx.doi.org/10.1140/epjc/s10052-019-6593-y}{\emph{Eur. Phys. J.
  C} {\bf 79} (2019) 83}, [\href{http://arxiv.org/abs/1811.01074}{{\tt
  1811.01074}}].

\bibitem{L3:2002hbp}
{\scshape L3} collaboration, P.~Achard et~al., \emph{{Search for single top
  production at LEP}},
  \href{http://dx.doi.org/10.1016/S0370-2693(02)02933-7}{\emph{Phys. Lett. B}
  {\bf 549} (2002) 290--300}, [\href{http://arxiv.org/abs/hep-ex/0210041}{{\tt
  hep-ex/0210041}}].

\bibitem{OPAL:2001spi}
{\scshape OPAL} collaboration, G.~Abbiendi et~al., \emph{{Search for single top
  quark production at LEP-2}},
  \href{http://dx.doi.org/10.1016/S0370-2693(01)01195-9}{\emph{Phys. Lett. B}
  {\bf 521} (2001) 181--194}, [\href{http://arxiv.org/abs/hep-ex/0110009}{{\tt
  hep-ex/0110009}}].

\bibitem{ALEPH:2002wad}
{\scshape ALEPH} collaboration, A.~Heister et~al., \emph{{Search for single top
  production in $e^{+} e^{-}$ collisions at $\sqrt{s}$ up to 209-GeV}},
  \href{http://dx.doi.org/10.1016/S0370-2693(02)02307-9}{\emph{Phys. Lett. B}
  {\bf 543} (2002) 173--182}, [\href{http://arxiv.org/abs/hep-ex/0206070}{{\tt
  hep-ex/0206070}}].

\bibitem{ZEUS:2003vfj}
{\scshape ZEUS} collaboration, S.~Chekanov et~al., \emph{{Search for single top
  production in ep collisions at HERA}},
  \href{http://dx.doi.org/10.1016/S0370-2693(03)00333-2}{\emph{Phys. Lett. B}
  {\bf 559} (2003) 153--170}, [\href{http://arxiv.org/abs/hep-ex/0302010}{{\tt
  hep-ex/0302010}}].

\bibitem{ZEUS:2011mya}
{\scshape ZEUS} collaboration, H.~Abramowicz et~al., \emph{{Search for
  single-top production in $ep$ collisions at HERA}},
  \href{http://dx.doi.org/10.1016/j.physletb.2012.01.025}{\emph{Phys. Lett. B}
  {\bf 708} (2012) 27--36}, [\href{http://arxiv.org/abs/1111.3901}{{\tt
  1111.3901}}].

\bibitem{Kikuchi:2000sv}
{\scshape CDF} collaboration, T.~Kikuchi, S.~K. Wolinski, L.~Demortier, S.~Kim
  and P.~Savard, \emph{{Search for Single Top Production with CDF}},
  {\emph{Int. J. Mod. Phys. A} {\bf 16S1A} (2001) 382--385}.

\bibitem{D0:2010dry}
{\scshape D0} collaboration, V.~M. Abazov et~al., \emph{{Search for Flavor
  Changing Neutral Currents via Quark-Gluon Couplings in Single Top Quark
  Production using 2.3 fb$^{-1}$ of $p\bar{p}$ Collisions}},
  \href{http://dx.doi.org/10.1016/j.physletb.2010.08.011}{\emph{Phys. Lett. B}
  {\bf 693} (2010) 81--87}, [\href{http://arxiv.org/abs/1006.3575}{{\tt
  1006.3575}}].

\bibitem{D0:2007wfn}
{\scshape D0} collaboration, V.~M. Abazov et~al., \emph{{Search for Production
  of Single Top Quarks Via Flavor-Changing Neutral Current Couplings}},
  \href{http://dx.doi.org/10.1103/PhysRevLett.99.191802}{\emph{Phys. Rev.
  Lett.} {\bf 99} (2007) 191802},
  [\href{http://arxiv.org/abs/hep-ex/0702005}{{\tt hep-ex/0702005}}].

\bibitem{ATLAS:2023ujo}
{\scshape ATLAS} collaboration, G.~Aad et~al., \emph{{Search for
  flavor-changing neutral $tqH$ interactions with $H\rightarrow \gamma\gamma$
  in $pp$ collisions at $\sqrt{s}$ = 13 TeV using the ATLAS detector}},
  \href{http://arxiv.org/abs/2309.12817}{{\tt 2309.12817}}.

\bibitem{ATLAS:2023qzr}
{\scshape ATLAS} collaboration, \emph{{Search for flavor-changing
  neutral-current couplings between the top quark and the $Z$ boson with LHC
  Run 2 proton-proton collisions at $\sqrt{s} = 13$ TeV with the ATLAS
  detector}},  \href{http://arxiv.org/abs/2301.11605}{{\tt 2301.11605}}.

\bibitem{ATLAS:2022gzn}
{\scshape ATLAS} collaboration, G.~Aad et~al., \emph{{Search for
  flavour-changing neutral current interactions of the top quark and the Higgs
  boson in events with a pair of \ensuremath{\tau}-leptons in pp collisions at
  $ \sqrt{s} $ = 13 TeV with the ATLAS detector}},
  \href{http://dx.doi.org/10.1007/JHEP06(2023)155}{\emph{JHEP} {\bf 2306}
  (2023) 155}, [\href{http://arxiv.org/abs/2208.11415}{{\tt 2208.11415}}].

\bibitem{ATLAS:2021amo}
{\scshape ATLAS} collaboration, G.~Aad et~al., \emph{{Search for
  flavour-changing neutral-current interactions of a top quark and a gluon in
  pp collisions at $\sqrt{s}=13$~TeV with the ATLAS detector}},
  \href{http://dx.doi.org/10.1140/epjc/s10052-022-10182-7}{\emph{Eur. Phys. J.
  C} {\bf 82} (2022) 334}, [\href{http://arxiv.org/abs/2112.01302}{{\tt
  2112.01302}}].

\bibitem{ATLAS:2018zsq}
{\scshape ATLAS} collaboration, M.~Aaboud et~al., \emph{{Search for
  flavour-changing neutral current top-quark decays $t\to qZ$ in proton-proton
  collisions at $\sqrt{s}=13$ TeV with the ATLAS detector}},
  \href{http://dx.doi.org/10.1007/JHEP07(2018)176}{\emph{JHEP} {\bf 07} (2018)
  176}, [\href{http://arxiv.org/abs/1803.09923}{{\tt 1803.09923}}].

\bibitem{ATLAS:2018jqi}
{\scshape ATLAS} collaboration, M.~Aaboud et~al., \emph{{Search for top-quark
  decays $t \to Hq$ with 36 fb$^{-1}$ of $pp$ collision data at $\sqrt{s}=13$
  TeV with the ATLAS detector}},
  \href{http://dx.doi.org/10.1007/JHEP05(2019)123}{\emph{JHEP} {\bf 05} (2019)
  123}, [\href{http://arxiv.org/abs/1812.11568}{{\tt 1812.11568}}].

\bibitem{ATLAS:2018xxe}
{\scshape ATLAS} collaboration, M.~Aaboud et~al., \emph{{Search for
  flavor-changing neutral currents in top quark decays $t\to Hc$ and $t \to Hu$
  in multilepton final states in proton-proton collisions at $\sqrt{s}= 13$ TeV
  with the ATLAS detector}},
  \href{http://dx.doi.org/10.1103/PhysRevD.98.032002}{\emph{Phys. Rev. D} {\bf
  98} (2018) 032002}, [\href{http://arxiv.org/abs/1805.03483}{{\tt
  1805.03483}}].

\bibitem{CMS:2016uzc}
{\scshape CMS} collaboration, V.~Khachatryan et~al., \emph{{Search for
  anomalous Wtb couplings and flavour-changing neutral currents in t-channel
  single top quark production in pp collisions at $\sqrt{s} =$ 7 and 8 TeV}},
  \href{http://dx.doi.org/10.1007/JHEP02(2017)028}{\emph{JHEP} {\bf 02} (2017)
  028}, [\href{http://arxiv.org/abs/1610.03545}{{\tt 1610.03545}}].

\bibitem{CMS:2017twu}
{\scshape CMS} collaboration, \emph{{Search for flavour changing neutral
  currents in top quark production and decays with three-lepton final state
  using the data collected at sqrt(s) = 13 TeV}},  CMS-PAS-TOP-17-017.

\bibitem{CMS:2017wcz}
{\scshape CMS} collaboration, A.~M. Sirunyan et~al., \emph{{Search for
  associated production of a Z boson with a single top quark and for tZ
  flavour-changing interactions in pp collisions at $ \sqrt{s}=8 $ TeV}},
  \href{http://dx.doi.org/10.1007/JHEP07(2017)003}{\emph{JHEP} {\bf 07} (2017)
  003}, [\href{http://arxiv.org/abs/1702.01404}{{\tt 1702.01404}}].

\bibitem{CMS:2021hug}
{\scshape CMS} collaboration, A.~Tumasyan et~al., \emph{{Search for
  Flavor-Changing Neutral Current Interactions of the Top Quark and Higgs Boson
  in Final States with Two Photons in Proton-Proton Collisions at
  $\sqrt{s}=13\text{ }\text{ }\mathrm{TeV}$}},
  \href{http://dx.doi.org/10.1103/PhysRevLett.129.032001}{\emph{Phys. Rev.
  Lett.} {\bf 129} (2022) 032001}, [\href{http://arxiv.org/abs/2111.02219}{{\tt
  2111.02219}}].

\bibitem{CMS:2023tir}
{\scshape CMS} collaboration, \emph{{Search for flavor-changing neutral current
  interactions of the top quark in final states with a photon and additional
  jets in proton-proton collisions at $\sqrt{s}$=13 TeV}},  CMS-PAS-TOP-21-013.

\bibitem{CMS:2020utv}
{\scshape CMS} collaboration, A.~M. Sirunyan et~al., \emph{{First measurement
  of the cross section for top quark pair production with additional charm jets
  using dileptonic final states in pp collisions at s=13TeV}},
  \href{http://dx.doi.org/10.1016/j.physletb.2021.136565}{\emph{Phys. Lett. B}
  {\bf 820} (2021) 136565}, [\href{http://arxiv.org/abs/2012.09225}{{\tt
  2012.09225}}].

\bibitem{Mandrik:2018yhe}
{\scshape FCC Study Group} collaboration, P.~Mandrik, \emph{{Prospect for top
  quark FCNC searches at the FCC-hh}},
  \href{http://dx.doi.org/10.1088/1742-6596/1390/1/012044}{\emph{J. Phys. Conf.
  Ser.} {\bf 1390} (2019) 012044}, [\href{http://arxiv.org/abs/1812.00902}{{\tt
  1812.00902}}].

\bibitem{Zarnecki:2018lup}
{\scshape CLICdp} collaboration, A.~F. Zarnecki, \emph{{Top-quark physics at
  the first CLIC stage}},
  \href{http://dx.doi.org/10.22323/1.340.0659}{\emph{PoS} {\bf ICHEP2018}
  (2019) 659}, [\href{http://arxiv.org/abs/1810.05487}{{\tt 1810.05487}}].

\bibitem{Cakir:2018ruj}
O.~Cakir, A.~Yilmaz, I.~Turk~Cakir, A.~Senol and H.~Denizli, \emph{{Probing top
  quark FCNC $tq\gamma$ and $tqZ$ couplings at future electron-proton
  colliders}},
  \href{http://dx.doi.org/10.1016/j.nuclphysb.2019.114640}{\emph{Nucl. Phys. B}
  {\bf 944} (2019) 114640}, [\href{http://arxiv.org/abs/1809.01923}{{\tt
  1809.01923}}].

\bibitem{CLICdp:2018esa}
{\scshape CLICdp} collaboration, H.~Abramowicz et~al., \emph{{Top-Quark Physics
  at the CLIC Electron-Positron Linear Collider}},
  \href{http://dx.doi.org/10.1007/JHEP11(2019)003}{\emph{JHEP} {\bf 11} (2019)
  003}, [\href{http://arxiv.org/abs/1807.02441}{{\tt 1807.02441}}].

\bibitem{TurkCakir:2017rvu}
I.~Turk~Cakir, A.~Yilmaz, H.~Denizli, A.~Senol, H.~Karadeniz and O.~Cakir,
  \emph{{Probing the Anomalous FCNC Couplings at Large Hadron Electron
  Collider}}, \href{http://dx.doi.org/10.1155/2017/1572053}{\emph{Adv. High
  Energy Phys.} {\bf 2017} (2017) 1572053},
  [\href{http://arxiv.org/abs/1705.05419}{{\tt 1705.05419}}].

\bibitem{Wang:2017pdg}
X.~Wang, H.~Sun and X.~Luo, \emph{{Searches for the Anomalous FCNC Top-Higgs
  Couplings with Polarized Electron Beam at the LHeC}},
  \href{http://dx.doi.org/10.1155/2017/4693213}{\emph{Adv. High Energy Phys.}
  {\bf 2017} (2017) 4693213}, [\href{http://arxiv.org/abs/1703.02691}{{\tt
  1703.02691}}].

\bibitem{Babu:1987kp}
K.~Babu, X.-G. He and E.~Ma, \emph{{New Supersymmetric Left-Right Gauge Model:
  Higgs Boson Structure and Neutral Current Analysis}},
  \href{http://dx.doi.org/10.1103/PhysRevD.36.878}{\emph{Phys.Rev.} {\bf D36}
  (1987) 878}.

\bibitem{Ma:2010us}
E.~Ma, \emph{{Dark Left-Right Model: CDMS, LHC, etc}},
  \href{http://dx.doi.org/10.1088/1742-6596/315/1/012006}{\emph{J.Phys.Conf.Ser.}
  {\bf 315} (2011) 012006}, [\href{http://arxiv.org/abs/1006.3804}{{\tt
  1006.3804}}].

\bibitem{Pati:1974yy}
J.~C. Pati and A.~Salam, \emph{{Lepton Number as the Fourth Color}},
  \href{http://dx.doi.org/10.1103/PhysRevD.10.275,
  10.1103/PhysRevD.11.703.2}{\emph{Phys. Rev.} {\bf D10} (1974) 275--289}.

\bibitem{Mohapatra:1974gc}
R.~Mohapatra and J.~C. Pati, \emph{{A Natural Left-Right Symmetry}},
  \href{http://dx.doi.org/10.1103/PhysRevD.11.2558}{\emph{Phys.Rev.} {\bf D11}
  (1975) 2558}.

\bibitem{Senjanovic:1975rk}
G.~Senjanovic and R.~N. Mohapatra, \emph{{Exact Left-Right Symmetry and
  Spontaneous Violation of Parity}},
  \href{http://dx.doi.org/10.1103/PhysRevD.12.1502}{\emph{Phys. Rev.} {\bf D12}
  (1975) 1502}.

\bibitem{Mohapatra:1977mj}
R.~N. Mohapatra, F.~E. Paige and D.~Sidhu, \emph{{Symmetry Breaking and
  Naturalness of Parity Conservation in Weak Neutral Currents in Left-Right
  Symmetric Gauge Theories}},
  \href{http://dx.doi.org/10.1103/PhysRevD.17.2462}{\emph{Phys.Rev.} {\bf D17}
  (1978) 2462}.

\bibitem{Ma:1986we}
E.~Ma, \emph{{Particle Dichotomy and Left-Right Decomposition of E(6)
  Superstring Models}},
  \href{http://dx.doi.org/10.1103/PhysRevD.36.274}{\emph{Phys. Rev.} {\bf D36}
  (1987) 274}.

\bibitem{Frank:2005rb}
M.~Frank, I.~Turan and M.~Sher, \emph{{Neutrino masses in the effective rank-5
  subgroups of E(6): Supersymmetric case}},
  \href{http://dx.doi.org/10.1103/PhysRevD.71.113002}{\emph{Phys. Rev. D} {\bf
  71} (2005) 113002}, [\href{http://arxiv.org/abs/hep-ph/0503084}{{\tt
  hep-ph/0503084}}].

\bibitem{Ashrythesis}
M.~Ashry, \emph{{TeV scale left-right symmetric model with minimal Higgs
  sector}},  {M.Sc. thesis}, U. Cairo, 2015.

\bibitem{Ashry:2013loa}
M.~Ashry and S.~Khalil, \emph{{Phenomenological aspects of a TeV-scale
  alternative left-right model}},
  \href{http://dx.doi.org/10.1103/PhysRevD.96.059901,
  10.1103/PhysRevD.91.015009}{\emph{Phys. Rev.} {\bf D91} (2015) 015009},
  [\href{http://arxiv.org/abs/1310.3315}{{\tt 1310.3315}}].

\bibitem{Bahrami:2016has}
S.~Bahrami, M.~Frank, D.~K. Ghosh, N.~Ghosh and I.~Saha, \emph{{Dark matter and
  collider studies in the left-right symmetric model with vectorlike leptons}},
  \href{http://dx.doi.org/10.1103/PhysRevD.95.095024}{\emph{Phys. Rev.} {\bf
  D95} (2017) 095024}, [\href{http://arxiv.org/abs/1612.06334}{{\tt
  1612.06334}}].

\bibitem{Frank:2019nid}
M.~Frank, B.~Fuks and O.~\"Ozdal, \emph{{Natural dark matter and light bosons
  with an alternative left-right symmetry}},
  \href{http://dx.doi.org/10.1007/JHEP04(2020)116}{\emph{JHEP} {\bf 04} (2020)
  116}, [\href{http://arxiv.org/abs/1911.12883}{{\tt 1911.12883}}].

\bibitem{Frank:2021ekj}
M.~Frank, C.~Majumdar, P.~Poulose, S.~Senapati and U.~A. Yajnik, \emph{{Vacuum
  structure of Alternative Left-Right Model}},
  \href{http://dx.doi.org/10.1007/JHEP03(2022)065}{\emph{JHEP} {\bf 03} (2022)
  065}, [\href{http://arxiv.org/abs/2111.08582}{{\tt 2111.08582}}].

\bibitem{Frank:2020odd}
M.~Frank, C.~Majumdar, P.~Poulose, S.~Senapati and U.~A. Yajnik,
  \emph{{Exploring $0\nu\beta\beta$ and leptogenesis in the alternative
  left-right model}},
  \href{http://dx.doi.org/10.1103/PhysRevD.102.075020}{\emph{Phys. Rev. D} {\bf
  102} (2020) 075020}, [\href{http://arxiv.org/abs/2008.12270}{{\tt
  2008.12270}}].

\bibitem{Frank:2022tbm}
M.~Frank, C.~Majumdar, P.~Poulose, S.~Senapati and U.~A. Yajnik, \emph{{Dark
  matter in the Alternative Left Right model}},
  \href{http://dx.doi.org/10.1007/JHEP12(2022)032}{\emph{JHEP} {\bf 12} (2022)
  032}, [\href{http://arxiv.org/abs/2211.04286}{{\tt 2211.04286}}].

\bibitem{Hahn:2010zi}
T.~Hahn, \emph{{Feynman Diagram Calculations with FeynArts, FormCalc, and
  LoopTools}}, \href{http://dx.doi.org/10.22323/1.093.0078}{\emph{PoS} {\bf
  ACAT2010} (2010) 078}, [\href{http://arxiv.org/abs/1006.2231}{{\tt
  1006.2231}}].

\bibitem{Hahn:2016ebn}
T.~Hahn, S.~Pa\ss{}ehr and C.~Schappacher, \emph{{FormCalc 9 and Extensions}},
  \href{http://dx.doi.org/10.1088/1742-6596/762/1/012065}{\emph{PoS} {\bf
  LL2016} (2016) 068}, [\href{http://arxiv.org/abs/1604.04611}{{\tt
  1604.04611}}].

\bibitem{Christensen:2009jx}
N.~D. Christensen, P.~de~Aquino, C.~Degrande, C.~Duhr, B.~Fuks, M.~Herquet
  et~al., \emph{{A Comprehensive approach to new physics simulations}},
  \href{http://dx.doi.org/10.1140/epjc/s10052-011-1541-5}{\emph{Eur. Phys. J.
  C} {\bf 71} (2011) 1541}, [\href{http://arxiv.org/abs/0906.2474}{{\tt
  0906.2474}}].

\bibitem{Alloul:2013bka}
A.~Alloul, N.~D. Christensen, C.~Degrande, C.~Duhr and B.~Fuks,
  \emph{{FeynRules 2.0 - A complete toolbox for tree-level phenomenology}},
  \href{http://dx.doi.org/10.1016/j.cpc.2014.04.012}{\emph{Comput. Phys.
  Commun.} {\bf 185} (2014) 2250--2300},
  [\href{http://arxiv.org/abs/1310.1921}{{\tt 1310.1921}}].

\bibitem{Dev:2016dja}
P.~S.~B. Dev, R.~N. Mohapatra and Y.~Zhang, \emph{{Probing the Higgs Sector of
  the Minimal Left-Right Symmetric Model at Future Hadron Colliders}},
  \href{http://dx.doi.org/10.1007/JHEP05(2016)174}{\emph{JHEP} {\bf 05} (2016)
  174}, [\href{http://arxiv.org/abs/1602.05947}{{\tt 1602.05947}}].

\bibitem{ParticleDataGroup:2022pth}
{\scshape Particle Data Group} collaboration, R.~L. Workman et~al.,
  \emph{{Review of Particle Physics}},
  \href{http://dx.doi.org/10.1093/ptep/ptac097}{\emph{PTEP} {\bf 2022} (2022)
  083C01}.

\bibitem{ATLAS:2019onj}
{\scshape ATLAS} collaboration, \emph{{Measurement of the top-quark decay width
  in top-quark pair events in the dilepton channel at $\sqrt{s}=13$ TeV with
  the ATLAS detector}},  ATLAS-CONF-2019-038.

\bibitem{CMS:2018ipm}
{\scshape CMS} collaboration, A.~M. Sirunyan et~al., \emph{{Search for
  high-mass resonances in dilepton final states in proton-proton collisions at
  13 TeV}}, \href{http://dx.doi.org/10.1007/JHEP06(2018)120}{\emph{JHEP} {\bf
  06} (2018) 120}, [\href{http://arxiv.org/abs/1803.06292}{{\tt 1803.06292}}].

\bibitem{ATLAS:2017fih}
{\scshape ATLAS} collaboration, M.~Aaboud et~al., \emph{{Search for new
  high-mass phenomena in the dilepton final state using 36 inverse fb of
  proton-proton collision data at 13 TeV with the ATLAS detector}},
  \href{http://dx.doi.org/10.1007/JHEP10(2017)182}{\emph{JHEP} {\bf 10} (2017)
  182}, [\href{http://arxiv.org/abs/1707.02424}{{\tt 1707.02424}}].

\bibitem{OPAL:1998znn}
{\scshape OPAL} collaboration, K.~Ackerstaff et~al., \emph{{Search for stable
  and longlived massive charged particles in e+ e- collisions at s**(1/2) =
  130-GeV - 183-GeV}},
  \href{http://dx.doi.org/10.1016/S0370-2693(98)00518-8}{\emph{Phys. Lett. B}
  {\bf 433} (1998) 195--208}, [\href{http://arxiv.org/abs/hep-ex/9803026}{{\tt
  hep-ex/9803026}}].

\bibitem{ATLAS:2016qxw}
{\scshape ATLAS} collaboration, \emph{{Expected sensitivity of ATLAS to FCNC
  top quark decays $t \rightarrow Zu$ and $t \rightarrow Hq$ at the High
  Luminosity LHC}},  ATL-PHYS-PUB-2016-019.

\bibitem{Cremer:2023gne}
L.~Cremer, J.~Erdmann, R.~Harnik, J.~L. Sp\"ah and E.~Stamou, \emph{{Leveraging
  on-shell interference to search for FCNCs of the top quark and the Z boson}},
  \href{http://dx.doi.org/10.1140/epjc/s10052-023-11982-1}{\emph{Eur. Phys. J.
  C} {\bf 83} (2023) 871}, [\href{http://arxiv.org/abs/2305.12172}{{\tt
  2305.12172}}].

\end{thebibliography}\endgroup

\end{document}